\PassOptionsToPackage{unicode}{hyperref}
\PassOptionsToPackage{hyphens}{url}
\documentclass[
]{article}
\usepackage{xcolor}
\usepackage{amsmath,amssymb}
\usepackage{iftex}
\ifPDFTeX
  \usepackage[T1]{fontenc}
  \usepackage[utf8]{inputenc}
  \usepackage{textcomp} 
\else 
  \usepackage{unicode-math} 
  \defaultfontfeatures{Scale=MatchLowercase}
  \defaultfontfeatures[\rmfamily]{Ligatures=TeX,Scale=1}
\fi
\usepackage{lmodern}
\ifPDFTeX\else
\fi
\IfFileExists{upquote.sty}{\usepackage{upquote}}{}
\IfFileExists{microtype.sty}{
  \usepackage[]{microtype}
  \UseMicrotypeSet[protrusion]{basicmath} 
}{}
\makeatletter
\@ifundefined{KOMAClassName}{
  \IfFileExists{parskip.sty}{%
    \usepackage{parskip}
  }{
    \setlength{\parindent}{0pt}
    \setlength{\parskip}{6pt plus 2pt minus 1pt}}
}{
  \KOMAoptions{parskip=half}}
\makeatother
\usepackage{longtable,booktabs,array}
\usepackage{calc} 
\usepackage{etoolbox}
\makeatletter
\patchcmd\longtable{\par}{\if@noskipsec\mbox{}\fi\par}{}{}
\makeatother
\IfFileExists{footnotehyper.sty}{\usepackage{footnotehyper}}{\usepackage{footnote}}
\makesavenoteenv{longtable}
\usepackage{graphicx}
\makeatletter
\newsavebox\pandoc@box
\newcommand*\pandocbounded[1]{
  \sbox\pandoc@box{#1}%
  \Gscale@div\@tempa{\textheight}{\dimexpr\ht\pandoc@box+\dp\pandoc@box\relax}%
  \Gscale@div\@tempb{\linewidth}{\wd\pandoc@box}%
  \ifdim\@tempb\p@<\@tempa\p@\let\@tempa\@tempb\fi
  \ifdim\@tempa\p@<\p@\scalebox{\@tempa}{\usebox\pandoc@box}%
  \else\usebox{\pandoc@box}%
  \fi%
}
\def\fps@figure{htbp}
\makeatother
\providecommand{\tightlist}{%
  \setlength{\itemsep}{0pt}\setlength{\parskip}{0pt}}
\usepackage{hyperref}
\usepackage{bookmark}
\IfFileExists{xurl.sty}{\usepackage{xurl}}{} 
\hypersetup{
  pdftitle={Registry Descriptions Go Stale Unevenly: An 89-Day Measurement of Model Context Protocol Drift{,} and Why Drift-Ranked Re-Auditing Under-Covers It},
  pdfauthor={Gautam Bharti, Independent Researcher, ORCID 0009-0001-4448-1438, gautamgb@gmail.com},
  hidelinks,
  pdfcreator={LaTeX via pandoc}}

\title{Registry Descriptions Go Stale Unevenly: An 89-Day Measurement of
Model Context Protocol Drift, and Why Drift-Ranked Re-Auditing
Under-Covers It}
\author{Gautam Bharti\\ Independent Researcher\\ ORCID 0009-0001-4448-1438 \quad \texttt{gautamgb@gmail.com}}
\date{Preprint, 2026}

\usepackage[letterpaper,margin=1in]{geometry}
\usepackage{etoolbox}
\AtBeginEnvironment{longtable}{\small}
\begin{document}
\maketitle

\subsection{Abstract}\label{abstract}

Security studies of the Model Context Protocol (MCP) ecosystem have
grown quickly, and they share a design: each audits a registry at a
single point in time. None reports how long a server's published
registry description - the field a registry-side screen reads - stays
current - a \emph{necessary} condition for any
description-level finding to still apply, though not a sufficient one:
we measure the shelf-life of the audited text, not the validity of a
security finding itself, which turns on live tool behavior we do not
observe (§7.1). We reconstruct 120 observations of the official MCP
registry over 88.6 days, covering 19,099 distinct servers as it grew
from 3,510 to 18,966. \textbf{Our central result is a policy one:} you
cannot keep description-level findings current by re-auditing the
servers that drift most. At a top-5\% re-audit budget, ranking by prior
drift catches only \textbf{\textasciitilde20\% of the previously-seen
servers whose description changes} in a held-out window - versus
\textasciitilde27\% for descriptor drift overall - and only
\textasciitilde10\% of \emph{all} description changers. The limit is not
that description drift is unpredictable - prior description change lifts
next-period probability \textbf{4.8x} (13.8\% vs 2.8\%) - but that the
signal is \emph{exhausted almost immediately}: only \textbf{5.0\% of the
population has any prior description-change history at all}, against
16.4\% for descriptors, so a top-5\% budget consumes the entire ranking
and caps near 20\%, and every slot beyond it is filled by tie-break
rather than by signal. Roughly half of \emph{all} description changes
then land on new arrivals a history ranking cannot reach by construction
(the highest-drift group). The control that fits is
content-binding - revalidate the moment a description\textquotesingle s
hash moves - plus a sized periodic full-catalog sweep for the new
arrivals and the tail; a drift-history ranking is at best a partial,
blind-to-new-arrivals control. This is scanner hygiene - keeping a
description-level auditor\textquotesingle s own findings current - not a
runtime trust signal for an agent about to invoke a tool.

The measurements behind this are the paper\textquotesingle s second
contribution. Description drift is concentrated and slow: of servers
observed across at least ten intervals, three-quarters never change, the
most active 5\% generate 61\% of all change \emph{events}, and directly
measured, \textbf{only 11.9\% of a cohort\textquotesingle s descriptors
change within 30 days} (three non-overlapping cohorts span 11.4--12.3\%;
§3.2). Naively compounding the daily change rate predicts 35.8\% at 30
days (73\% at 89) - a \textasciitilde3.0x overestimate at 30 days that
widens to \textasciitilde3.8x by 89 days, which we use only as a
heavy-tail \emph{diagnostic}, not a discovery; longer horizons rest on
progressively fewer independent cohorts, so we report 19.1\%-at-89-days
as a trend, not a headline. We scope every claim precisely: this is a
revalidation policy for the registry description text a
description-level screen reads, and bounds neither tool-poisoning
shelf-life nor the validity of any security finding - the live tool list
and input schemas an attacker manipulates are a separate surface this
data does not observe (§7.1). We release the panel, the figure
generator, and the analysis code. (A deployment observation on our own
scanner, §6, is secondary, single-deployment, and not among the
paper\textquotesingle s contributions.)

\begin{center}\rule{0.5\linewidth}{0.5pt}\end{center}

\pandocbounded{\includegraphics[keepaspectratio,alt={Registry drift measured three ways: drift-ranked re-auditing under-covers description revalidation; the audited text turns over far slower than naive compounding; change is heavily concentrated.}]{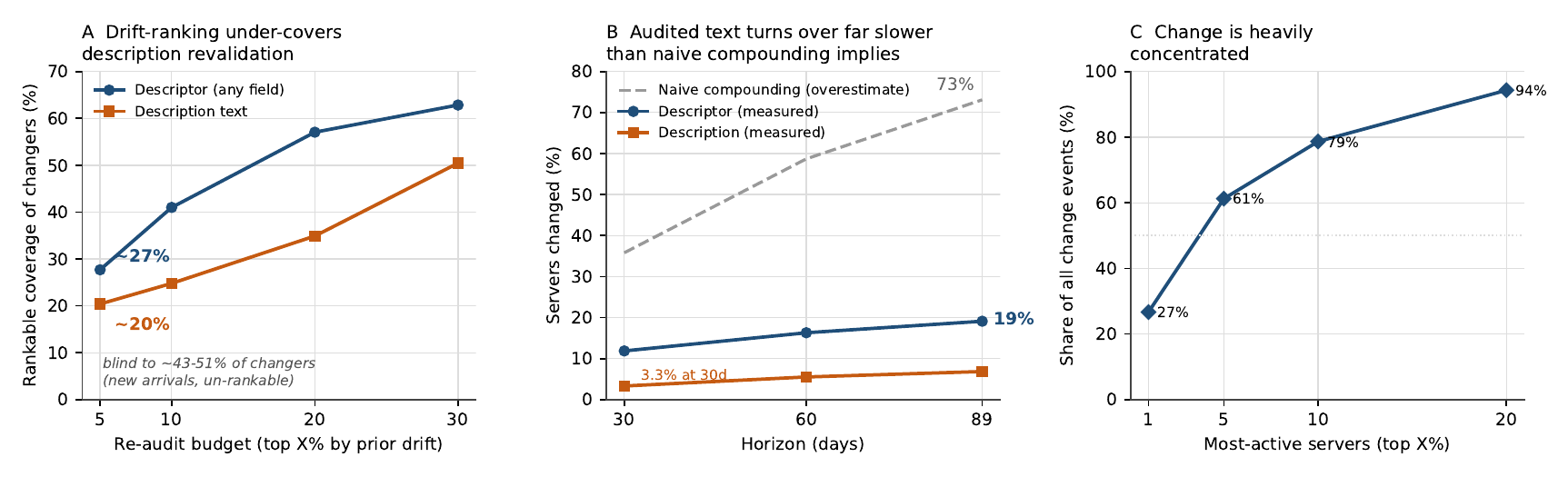}}

\textbf{Figure 1. Three findings, one panel each.} \textbf{(A)} Ranking
servers by prior drift and re-auditing the top X\% under-covers the
description-revalidation surface: at a top-5\% budget it catches
\textasciitilde27\% of previously-seen descriptor changers but only
\textasciitilde20\% of description changers, and is blind by
construction to the \textasciitilde43-51\% of changers that are new
arrivals it has never seen. \textbf{(B)} The audited text turns over
far slower than naive compounding of the daily rate predicts (89-day
descriptor survival 19\% measured vs 73\% compounded); description text,
the surface a description-level screen must recheck, changes slower still
- 3.3\% at 30 days. \textbf{(C)} Change is heavily concentrated: the most active 5\%
of servers produce 61\% of all change events. Every value is read from
the deposited \texttt{figures.json} and regenerated by
\texttt{make\_paper\_figure.py}. CC-BY-4.0.

\begin{center}\rule{0.5\linewidth}{0.5pt}\end{center}

\subsection*{Changes in v2}

This version corrects five claims in v1, found by a structured
re-review of the paper against its own deposited artifact. Each
correction is regenerable from the deposit; the analysis code that
produces them ships with the dataset.

\begin{itemize}
\tightlist
\item
  \textbf{§3.2, moving-block bootstrap.} v1 derived the block length
  from the observation series, but the bootstrap resamples cohort
  \emph{starts}: 40 starts span a median 42.0 days (recomputable from the
  panel's observation timestamps), not the 30 claimed.
  The block matching the horizon is 29, and the interval moves from
  {[}11.6, 12.2{]}\% to {[}11.5, 12.4{]}\%. v1 attributed the narrowness
  solely to cohort overlap; block length is also a lever, and the width
  profile now ships so that is checkable.
\item
  \textbf{§3.3/§5.3, sizing of the recommended control.} v1 converted a
  cohort survival share into a revalidation rate, which excludes servers
  arriving mid-window and collapses repeat changes. Measured directly as
  events on the live catalogue the load is 38.7/day over the trailing 30
  days (2.04 per 1,000 entries/day), not the \ensuremath{\sim}20/day v1
  reported - a \ensuremath{\sim}2x understatement, and the number an
  operator would budget on.
\item
  \textbf{§4.2, age and concentration.} v1 reported the two as close to
  independent axes. That was an artifact of ranking by cumulative change
  count, which confounds hazard with exposure. Under exposure
  normalisation the top 5\% is 89.5\% newborn against an 81.3\% baseline.
  The policy conclusion is unchanged: evaluated as a targeting arm, the
  exposure-normalised ranking does not improve description coverage.
\item
  \textbf{§4.2, the age-hazard gradient.} v1 reported the decline as a
  single pooled ratio (5.5x any-field, 10x description-only). The age-week
  buckets are near-disjoint arrival cohorts - week-ten exposure requires 70
  days of follow-up - so the pooled magnitude is inflated by composition. On
  one cohort observable across every age week the decline is
  20.2 \ensuremath{\rightarrow} 6.9, and v2 states a bracket rather than a
  point. The direction is unchanged.
\item
  \textbf{§4.3, coverage versus text-currency.} v1 claimed an audit's
  coverage claim decays faster than its findings. It cannot: the
  self-reported count is parsed from the description text, so all 265
  count changes coincide with a description-hash change. The corrected
  claim is about magnitude, not rate.
\end{itemize}

Consequential edits follow from those five. §2.4 and §6.2 restate the
panel-side cross-check as 34.6/day over the deployment's own window, and
§2.4 now rests the survival measurement on the three non-overlapping
cohorts rather than the bootstrap interval, which §3.2 no longer reads as
corroboration. §1 and Figure 1B drop "findings decay" for "the audited text
turns over": the panel measures text currency, not the validity of
findings (§7.1). Two §5.3 marginal-return figures (+0.72, +2.62) are
restated as +0.71 and +2.63, read from the addenda at full precision rather
than differenced from the rounded table; §5.3 also now states that the
marginal-return shape above the 5\% budget is a named-panel result that
reverses under the released panel's tokenization, because those slots carry
no drift signal. §5.1's "ten times the description
hazard" becomes the §4.2 bracket. Two section headings were rewritten to
match their corrected bodies (§4.2, §4.3), the Figure 1 caption gains the
30-day description figure, and §5.4 now states that its counts come from
the unreleased verdict store. Figure 1's plotted values are unchanged;
its panel-A legend and annotation styling moved, and its PDF metadata is
now stripped for byte-reproducibility. The concentration of drift, the measured survival
curves, and the finding that drift-ranked re-auditing under-covers
description revalidation are unchanged; what moved is the sizing of the
recommended control, the uncertainty treatment in §3.2, and two mechanism
claims in §4.2-§4.3.

\subsection{1. Introduction}\label{1-introduction}

The Model Context Protocol gives agentic systems a uniform way to
discover and call third-party tools, and its public registries have
become, in under a year, a software supply chain with tens of thousands
of entries. A security literature has formed around it at speed:
large-scale registry audits {[}1{]}, health-and-maintainability studies
{[}2{]}, attack analyses over harvested tool corpora {[}3{]}, and
client-side threat modeling of tool poisoning {[}4{]}. These works
differ in scope and method but share one structural property: each
observes the ecosystem once. The audit is a photograph.

A photograph of a supply chain has a shelf life, and for MCP nobody has
measured it. The question matters operationally - an integrator who
consumed any point-in-time registry audit, such as {[}1{]}, and acted on a
finding computed over a server's published description needs to know whether
the \emph{text those findings were computed over} is still the text
being served in September (a necessary condition for the finding to
still apply, though not a sufficient one - the live tool behavior can
change under a stable description, and vice versa). To be unambiguous:
this is a data-currency question for the auditor, not a runtime
trust-TTL for an agent about to call a tool - nothing here licenses
"this server was audited N days ago, so it is safe to invoke now." And
it matters methodologically, because every claim of the form "X\% of
descriptions exhibit property P" silently expires at a rate the
literature has not characterized.

This paper measures that rate, and finds that the interesting fact is
not its magnitude but its shape. The two findings below are load-bearing
and deposit-backed; a deployment observation on our own scanner (§6) is
secondary and, being a single unreleased store, not independently
replicable - we keep it out of the contribution claims.

\textbf{Drift is concentrated, and an audit ages slowly and measurably}
(§3--4). Three quarters of registry entries (of those observed across
\ensuremath{\geq}10 intervals) never change over 88.6 days; the most active twentieth
produces three fifths of all change events. Directly measured, only
11.9\% of a cohort\textquotesingle s descriptors change within 30 days -
the three non-overlapping 30-day cohorts give 11.4 / 11.9 / 12.3\%,
which we take as the primary uncertainty; a moving-block bootstrap over
overlapping cohorts gives a 95\% CI of (11.5--12.4) that is no wider than
that spread and excludes its low end, so it understates between-cohort
variance rather than corroborating it - so
the audited text turns over far slower than a naive rate
implies. The naive rate is
not merely imprecise: compounding the daily change rate predicts 35.8\%
at 30 days (and 73\% at 89), because compounding an event rate into a
population share assumes each event lands on a fresh server - the
textbook heavy-tail failure, which we use as a diagnostic (§3.2) rather
than a discovery. We report the 30-day figure as the robust one and
treat the 89-day survival (19.1\% across nine overlapping cohorts; the one
non-overlapping cohort gives 21.9\%) as a weak
trend.

\textbf{Concentration is prospectively exploitable - but not on the
surface a screen must revalidate} (§5). At a top-5\% re-audit budget,
ranking by prior drift catches \textasciitilde27\% of the
\emph{previously-seen} servers that change (a 5--6x lift over random;
\textasciitilde13--17\% of all changers, since half are new arrivals it
cannot rank; 47\% of change events); the \emph{same} budget catches only
\textasciitilde20\% of the previously-seen servers whose
\emph{description} changes, the surface a description-level screen
actually has to recheck. The limit is not that description rewrites are
unpredictable - prior description change lifts next-period probability
4.8x - but that only ~5\% of servers carry any prior description change to
rank on, so a history ranking exhausts its signal pool almost immediately. The operational consequence is specific and,
we think, the paper\textquotesingle s most useful result \textbf{for
operators of description-level registry screens}: \textbf{history-ranked
re-auditing is the wrong control for description revalidation} - the
right one is content-binding (revalidate when the description hash
moves) plus a periodic calendar sweep, neither of which a churn ranking
provides. We scope this to description revalidation deliberately: the
surface is the registry description text, not the live tool list or
input schema an attacker ultimately manipulates (§7.1), which our data
does not observe.

Separately, §6 reports a deployment observation on our own scanner:
applying the same lens to its verdict store surfaces a second staleness
producer - the scanner\textquotesingle s input pipeline lagging the
registry, which mints verdicts \emph{wrong at birth} rather than merely
aged. We include it because instrument lag is absent from the threat
models we know, but we do not count it among the paper\textquotesingle s
contributions: the store is a single unreleased deployment and
§6\textquotesingle s counts are not independently replicable (§7).

We release the panel, the figure generator, and the analysis code; §6
also notes the three production changes the deployment observation
forced.

\subsection{2. Data and Method}\label{2-data-and-method}

\subsubsection{2.1 The registry panel}\label{21-the-registry-panel}

Throughout, \emph{description} means the registry record's own
\texttt{description} field - a blurb the registry caps at 100 characters -
and not the per-tool descriptions returned by a server's \texttt{tools/list}
response, which this panel never observes and which are the surface a
tool-poisoning threat model concerns (§7.1).

The official MCP registry is synced every four hours by our production
pipeline; each successful sync commits the full snapshot to version
control when it changed. The dated per-day snapshot files are ephemeral
(they die with the CI runner), but the committed history is a faithful
longitudinal record: \textbf{120 revisions from 2026-04-30 to
2026-07-28} (88.6 days, up to six observations per day), during which
the corpus grew from 3,510 to 18,966 active servers; 19,099 distinct
server names appear.

We replay that history into a delta-encoded panel. For each observation
and server we record \texttt{f} = SHA-256 of the canonicalized server
descriptor (any-field drift) and \texttt{d} = SHA-256 of the description
alone (the description-revalidation surface), plus version and a parsed
self-reported tool count where present. Because 75.2\% of servers never
change, delta encoding compresses 51 MB of per-revision digests to 1.3
MB. The panel is a deterministic function of the public snapshot
history, pinned to snapshot commit \texttt{e1d19d3} (see Data and Code
Availability), so a replicator rebuilds it from version control and
confirms an identical file. The released (pseudonymized) panel
reproduces the named panel\textquotesingle s full finding signature -
every concentration tier, the 30/60/89-day survival curves, and an
identical observation-timestamp hash - asserted at
build time against the un-pseudonymized panel. A replicator can confirm the
released file against \texttt{SHA256SUMS} and re-derive every finding with
the included \texttt{paper\_figures.py}.

\subsubsection{2.2 The verdict panel}\label{22-the-verdict-panel}

The same repository\textquotesingle s history yields 108 observations
(2026-05-30 to 2026-07-28) of a production verdict store - a
point-in-time LLM screen over registry descriptions, serving advisory
verdicts on a public trust surface. Each verdict carries
\texttt{evaluated\_at} and, after 2026-07-19, \texttt{content\_hash} =
SHA-256 of the exact description judged. §6 joins this store to the
registry panel.

\subsubsection{2.3 Two traps for
replicators}\label{23-two-traps-for-replicators}

\textbf{\texttt{publishedAt} is per-version, not per-server.} The
registry rewrites \texttt{publishedAt} on every version publish -
verified directly: of 225 servers that changed version across one
snapshot pair, 225 also changed \texttt{publishedAt}. Any survival
analysis anchored on that field measures nothing. Our first attempt did
exactly this and produced a plausible, internally-consistent, wrong
result; we caught it only because an independently derived quantity
contradicted it.

\textbf{Hash derivations must mirror the writer byte-for-byte.} The
store\textquotesingle s writer strips whitespace before hashing; an
offline re-derivation that hashed raw registry text produced 34 false
staleness accusations out of 781 (§6.1). The general rule: before
comparing an independently derived hash to a stored one, reproduce the
writer\textquotesingle s exact normalization - or call the
writer\textquotesingle s own code.

\subsubsection{2.4 Estimator
cross-validation}\label{24-estimator-cross-validation}

We compute rates more than one way and report where the estimates
diverge as readily as where they agree - in this paper the divergence
\emph{is} the result. One agreement is load-bearing: (i) the per-day
event \emph{rate} - the direct interval-diff rate and the
registry\textquotesingle s \texttt{publishedAt}-implied version-publish
intensity accumulate to the same \textasciitilde36\% of servers by 30
days (§3.2). A second, weaker consistency check (not load-bearing) is
(ii) the post-judgment drift rate: verdict-dating gives
\textasciitilde26/day and the panel\textquotesingle s independent
description-change events 34.6/day over the same period
(§6.2) - the same order of magnitude, which is all we claim of it; we do
not treat 26 vs 34.6 as a precise match. The load-bearing \emph{disagreement} is between those
event-intensity estimators and direct cohort survival: compounding the
event rate - and, separately, the version-recency population share -
both put \textasciitilde70\% of servers changed by \textasciitilde90
days, while direct survival of the cohort present at start measures
\textasciitilde19\% (§3.2). That two independent routes to event
intensity overshoot direct survival by the same \textasciitilde3.8x is
what isolates the error to the population-share conversion, not the rate
and not the survival measurement - the latter\textquotesingle s own
reliability resting on the three
non-overlapping cohorts (§3.2), not a
second estimator.

\subsection{3. How Fast Does the Registry
Change?}\label{3-how-fast-does-the-registry-change}

\subsubsection{3.1 The rate}\label{31-the-rate}

Across 119 inter-observation intervals, the share of common servers
whose descriptor changed, normalized per day: median \textbf{1.47\%/day}
(mean 1.73; p10 1.01; p90 2.37). The mean is inflated by short-gap
intervals - a small absolute change divided by a fraction of a day - so
we use the median throughout. Description-only change is several times
rarer (§3.3).

\subsubsection{3.2 The compounding
error}\label{32-the-compounding-error}

It is tempting to convert that rate into a shelf life: 1 \ensuremath{-} (1 \ensuremath{-}
0.01465)\^{}N (the unrounded median of §3.1) gives 35.8\% of servers
changed by day 30, 58.7\% by day 60, 73.1\% by day 89 - an audit
"half-life" of about 47 days. Direct measurement refutes every one of
those numbers:

{\def\LTcaptype{none} 
\begin{longtable}[]{@{}llll@{}}
\toprule\noalign{}
Horizon & Compounded prediction & Measured (cohort mean) & Cohorts \\
\midrule\noalign{}
\endhead
\bottomrule\noalign{}
\endlastfoot
30 days & 35.8\% & \textbf{11.87\%} (range 10.37--13.71) & 60 \\
60 days & 58.7\% & \textbf{16.30\%} & 36 \\
89 days & 73.1\% & \textbf{19.11\%} & 9 \\
\end{longtable}
}

\textbf{Estimand.} Under commit-on-change sampling (an observation
exists only when the snapshot changed), we count a server as "changed
within N days" if its descriptor hash differs from its cohort-start
value at \emph{any} observation within the window - the
\emph{ever-flipped} estimand, which upper-bounds an endpoint-only
definition (a change that later reverts still counts). A cohort start is
each observation; a server is in the cohort if present at the start and
observed with \ensuremath{\geq}0.9N days of follow-up.

The 30-day figure is the most defensible single number in this paper,
but its uncertainty must be stated honestly: the 60 cohort starts
\emph{overlap} - each shares its 30-day future window with its
neighbours - so their range (10.4--13.7\%) is pseudo-replication, not
independent scatter. We therefore lead with the three genuinely
non-overlapping 30-day cohorts (starts \ensuremath{\geq}30 days apart): \textbf{11.4 /
11.9 / 12.3\%}, and treat that scatter as the honest uncertainty. A
moving-block bootstrap gives a 95\% CI of \textbf{{[}11.5, 12.4{]}\%}
(full recipe in \texttt{figures.json}: B = 2000 resamples, block length
= 29 cohort \emph{starts}, whose median calendar span is 29.1 days \ensuremath{\approx} the
30-day overlap horizon, seed 42, percentile method), but that interval is
no wider than the three-cohort scatter and excludes the 11.4\% cohort - because the bootstrap resamples overlapping cohorts that share
observations, it understates between-cohort variance rather than
confirming the low end, so we do not read it as corroboration. Block
length is a real lever on that width, and we state it rather than wave it
away: with only 60 cohort starts, a block of 29 is 48\% of the series, so
each resample is roughly two blocks in their original order and the
interval is mechanically tight; width is not monotone in the block length
but peaks near 20--25 starts and then collapses toward zero as the block
approaches \emph{n} (\texttt{figures.json} ships
\texttt{block\_bootstrap\_profile}: the interval at ten block lengths).
Even at its widest - block 20, lower
bound 11.39\% - the interval only reaches the lowest non-overlapping cohort
(11.38\%) to within 0.01 pp, a margin that flips with the bootstrap seed; no
block length puts that cohort comfortably inside. That is why the three
non-overlapping cohorts, not the CI, carry the uncertainty claim. The point estimate is stable in magnitude across a
3.5k\ensuremath{\rightarrow}14k corpus, though not flat: across the 60 cohort
starts it drifts up by roughly 0.9 pp (first-half mean 11.4\%, second-half
12.3\%), and the three non-overlapping cohorts run 12.3
\ensuremath{\rightarrow} 11.9 \ensuremath{\rightarrow} 11.4 in calendar
order. That drift is the same size as the spread we quote as the honest
uncertainty, so the spread is partly systematic rather than pure scatter. The 89-day figure, by contrast, admits only one
non-overlapping cohort and rests on nine overlapping early ones; we
treat it as a weak trend, not a measurement.

The error is structural, not sampling noise. The daily rate counts
\emph{events}; compounding it assumes each event strikes a fresh server.
On this population the assumption fails by design (§4), and the failure
generalizes: registries, package indexes, and user-activity streams are
typically heavy-tailed, so the compounding shortcut is biased upward
almost everywhere it is used. The cheap detector: compare distinct
entities that ever fired an event against total events. Here, 18,748
servers observed across \ensuremath{\geq}10 intervals produced 15,805 change events from
only 4,652 distinct servers.

A second, independent estimator confirms the event \emph{rate} - not the
survival. The share of servers whose \emph{current version} was
published within the last N days (from the registry\textquotesingle s
public \texttt{publishedAt}, no diffing) gives 36.7\% / 56.5\% / 70.4\%
at 30/60/90 days. These track the \emph{compounded} prediction (35.8 /
58.7 / 73.1), not the direct survival (11.9 / 16.3 / 19.1) - as they
must: version-recency counts new arrivals and per-version
\texttt{publishedAt} rewrites (§2.3), so like compounding it reflects
event intensity, not the survival of a fixed cohort. That an empirical
intensity estimator and an arithmetic one overshoot direct survival
together - by a similar factor (\textasciitilde3.7--3.8x), growing from \textasciitilde3.0x at 30 days to
\textasciitilde3.8x at 89 days - is exactly what localizes the error to
the population-share conversion rather than to the event rate or the
survival measurement. (This estimator reads the registry snapshot, not
the released panel, which carries no \texttt{publishedAt}.)

\subsubsection{3.3 Description drift, the revalidation
surface}\label{33-description-drift-the-revalidation-surface}

Measured directly, description-only change: \textbf{3.32\%} of servers
by 30 days (range 2.58--4.17 across 60 cohorts), 5.52\% by 60, 6.85\% by
89. That 3.32\% is a \emph{cohort survival} share - the fraction of servers
present at a cohort start that ever flip - so it is the wrong quantity to
size a control with: it excludes changes on servers that arrive mid-window
and collapses repeat changes to one. Measured directly as events on the
live catalog, the revalidation load is \textbf{38.7 description-hash
mismatches per day over the trailing 30 days (2.04 per 1,000 of the 18,966
entries live at cutoff, or 2.37 per 1,000 against the 16.3k mean corpus
over that window), and 28.4/day averaged over the whole window}, when the
corpus averaged about two-thirds of its final size (time-weighted mean
12.4k against 19.0k at cutoff; per entry the two agree at 2.3 and 2.4
mismatches per 1,000 entries per day, so the gap is corpus growth rather
than a change in per-entry churn) (\texttt{figures.json}:
\texttt{revalidation\_load}). Each one is a mutation of exactly the text a
description-level screen judged, and therefore must revalidate. Converting
the survival share instead would have given \ensuremath{\sim}20/day - a
\ensuremath{\sim}2x understatement, produced by precisely the new-arrival
blind spot we diagnose for \emph{ranking} in §4.2 and §5.2. We call this the
\emph{description-revalidation surface} deliberately, not "the
tool-poisoning surface": the registry description is the input a
description-level screen sees, but the behavior an attacker manipulates
lives in the live tool list and input schemas, which the registry does
not carry and this dataset does not observe (§7.1). The two surfaces can
move independently; our claims are about the one we measure.

\subsection{4. Who Changes?}\label{4-who-changes}

\subsubsection{4.1 A stable majority, a busy
minority}\label{41-a-stable-majority-a-busy-minority}

Of 18,748 servers observed across at least 10 intervals, \textbf{14,096
(75.2\%) recorded zero descriptor changes} in the whole window. The
change-count histogram: 0 \ensuremath{\rightarrow} 14,096 · 1 \ensuremath{\rightarrow} 2,183 · 2--4 \ensuremath{\rightarrow} 1,594 · 5--9 \ensuremath{\rightarrow}
525 · 10+ \ensuremath{\rightarrow} 350. "Never changes" here means the \emph{registry
descriptor} is stable; a server whose live tool behavior mutates under a
fixed description sits in this 75\% and is precisely the blind spot §7.1
names. Descriptor stability is not tool-behavior stability, and this
stable majority must not be read as a majority that is safe to leave
un-probed.

Concentration of the 15,805 change events: top 1\% of servers = 26.7\%,
top 5\% = 61.2\%, top 10\% = 78.7\%, top 20\% = 94.3\%. Publisher-level
concentration is milder but real: the top ten publishers account for
17.7\% of events (the single busiest, 5.5\%); 3,054 of 11,900 publishers
ever produced a change event. (Publisher-level figures need the
plaintext name and so are computed from the internal named panel, not
the pseudonymized release - see §7.)

The \ensuremath{\geq}10-interval eligibility filter - which by construction excludes
late arrivals, the high-hazard newborns of §4.2 - does not carry these
numbers. Varying the threshold from no filter (\ensuremath{\geq}1) through \ensuremath{\geq}20 moves
top-5\% event share only within 61.2--61.5\% and never-changed within
75.2--75.4\% (never-changed: 75.4 / 75.3 / 75.2 / 75.2 at \ensuremath{\geq}1/5/10/20;
top-5\%: 61.5 / 61.3 / 61.2 / 61.4). The filter removes
\textasciitilde300 short-window servers without shifting either
headline, so the concentration finding is not an artifact of censoring
the newborns.

\subsubsection{4.2 Drift is front-loaded, and count-ranking hides
it}\label{42-drift-is-front-loaded-and-count-ranking-hides-it}

Restricting to the servers first observed \emph{inside} the window (so age
is observed, not left-censored), the any-field hazard falls from
\textbf{38.0 changes per 1,000 server-days in a server\textquotesingle s
first week to 6.9 by week ten}; the description-only hazard falls
7.0 \ensuremath{\rightarrow} 0.7. Those pooled buckets are not the same
servers, and we do not read a ratio off them: week-ten exposure exists only
for servers that arrived early enough to be followed 70 days, while the
week-zero bucket is dominated by late arrivals. Recomputing the identical
estimator on the one arrival cohort observable across every age week
(\emph{n} = 5,594) gives 20.2 \ensuremath{\rightarrow} 6.9 and
3.6 \ensuremath{\rightarrow} 0.7. We therefore state the decline as a
\textbf{bracket - roughly 3x to 5.5x on the any-field surface and 5x to 10x
on description-only, depending on whether composition is held fixed} -
rather than a single pooled ratio. The direction is robust to every
decomposition we ran; the magnitude is not. New servers are where drift
lives.

Ranked by cumulative change count - the ranking §5 evaluates - the top 5\%
is \textbf{80.8\% newborn against an 81.3\% baseline}, apparently
indistinguishable. That near-equality is an artifact of the estimator, not
a property of the population. A cumulative count confounds hazard with
exposure: the count-ranked top 5\% is observed across 94.6 intervals
against an eligible mean of 88.5, so ranking by raw count systematically
under-selects the youngest servers and cancels their hazard excess.
Re-ranking the identical 18,748-server population by changes \emph{per
observed interval} puts the top 5\% at \textbf{89.5\% newborn, +8.2 pp over
baseline}, and the enrichment holds from +6.5 to +8.3 pp as the
eligibility floor is raised from 10 to 60 observed intervals. Age and
concentration are independent only under the \emph{count} definition of
concentration used in §4.1 and §5; under an exposure-normalised one they
are not, and a count-ranked policy simply cannot see the enrichment. The
correction changes the mechanism, not the policy: evaluated as a targeting
arm under the identical §5 protocol, the exposure-normalised ranking does
not rescue description coverage (19.3\% against 19.3\% at the 44-day cut;
21.3\% against 21.5\% at 60 days), because the limit in §5.3 is how little
description history exists at all, not how it is sorted.

\subsubsection{4.3 Self-reported tool counts almost only
grow}\label{43-self-reported-tool-counts-almost-only-grow}

The registry caps the description at 100 characters (across all 18,966
entries in the pinned snapshot the maximum is exactly 100 and the median
87), so it is a short blurb rather than documentation - which bounds
both what a description-level screen can judge and what a self-reported
count can mean. 861 servers self-report a tool count in their
description. Across the
window: 248 increase events against 17 decreases (93.6\% increases), net
\textbf{+2,435 tools}. Single-day jumps include 8 \ensuremath{\rightarrow} 27 tools.
Self-reported counts are prose, not a probe of the real tool list
(Limitation §7.2), and they cannot decay faster than the text: a stated
count lives \emph{inside} the description, so every one of the 265 count
changes coincides with a description-hash change - the count signal is a
strict subset of the text-currency signal, by construction. What the counts
add is the \emph{magnitude} of a claimed inventory change, not a rate, and
we treat it as a secondary observation: it does not bear on the
content-binding recommendation, which triggers on the hash regardless. Most edits move the count a little - the
median change is 1.11x - but the tail is real: 38 of the 265 changes are
\ensuremath{\geq}1.5x and 14 are \ensuremath{\geq}2x, up to 4.7x. So the size of a claimed
inventory change and the fact of a text change do not move in proportion,
and an auditor who tracks only \emph{whether} the text changed learns
nothing about how large a change the publisher is claiming.

\subsubsection{4.4 Delisting is negligible, but reappearance is
not}\label{44-delisting-is-negligible}

885 servers vanished from at least one observation; only 133 (0.7\% of
the final corpus) were absent at the end. Removal is not a meaningful
decay channel; mutation is.

The complement is more interesting than the headline. Because almost
nothing leaves permanently, almost everything that vanishes
\emph{returns}: we record \textbf{778 reappearance events across 764
distinct names}, 14 of which vanished and returned more than once. A
registry name is the referent an allowlist, a catalog entry, or a
procurement approval keys on, so a name that leaves and comes back is
a structural analogue of the package-reclamation pattern
known from npm and PyPI - and unlike ordinary mutation, it is invisible to
a consumer who only checks whether the name is still listed. The
registry's namespacing weakens that analogy: every name takes the form
\texttt{namespace/name} - 71.8\% of them under \texttt{io.github.*},
which the registry binds to the corresponding GitHub account, the rest
under DNS-verified domains - so a name returning to a \emph{different}
party would require control of the same authenticated namespace, not
merely an unclaimed string. The panel carries content hashes, not
ownership, so it cannot observe whether that occurred, and we do not
claim it did.

We can bound how often the returned artifact differs from the one that
left. Of the 778 reappearances, \textbf{12 (1.5\%) returned with a
different description hash} and \textbf{43 (5.5\%) with a different
descriptor hash}. The overwhelming majority return byte-identical, which
is consistent with transient sync or registry-side availability blips
rather than re-registration. So the reappearance channel is real,
enumerable, and \emph{small}; we report it because a null of this shape
is what a threat model needs in order to rank a plausible
attack path, and because no prior MCP registry study reports it at all.
We stress the limit: the panel carries content hashes, not ownership or
publisher identity, so these 12 cases establish that content changed
across a disappearance, not that the name changed hands.

\subsection{5. Can Staleness Be
Predicted?}\label{5-can-staleness-be-predicted}

\subsubsection{5.1 Retrospective lift}\label{51-retrospective-lift}

Splitting the window at its midpoint (2026-06-29; 3,492 servers present
in both halves): P(drift in H2 \textbar{} drifted in H1) =
\textbf{31.2\%}, against 4.4\% for servers stable in H1 - a \textbf{7.1x
lift}. On description text alone the same split gives \textbf{13.8\%}
against 2.8\%, a \textbf{4.8x lift}. Description rewrites are therefore
substantially predictable from prior description rewrites; §5.3 shows
why that predictability nonetheless buys little coverage.

Two properties of this estimate constrain how far it generalises. First,
the base is servers present at \emph{both} the first observation and the
midpoint, which is 99.5\% of the panel-start corpus and only
\textbf{18.4\% of the corpus at the end}. The conditional therefore
describes the oldest tier and excludes by construction the new arrivals
that carry roughly 5--10x the description hazard (§4.2) and about
half of all description changes (§5.2). Second, it is one split of one
window; we report it as a retrospective description of this corpus, not
as a forecast.

\subsubsection{5.2 Prospective
targeting}\label{52-prospective-targeting}

Retrospective concentration is not a policy, so we test the policy: rank
servers by drift observed in a training window only, re-audit the top
X\%, and measure the share of \emph{held-out} change events captured. We
use two training cut-dates (44-day and 60-day windows from panel start);
with only two cut-points we report the spread across them as a
sensitivity range, not a confidence interval:

We report entity coverage, not event share, because §§3--4 warn that
events \ensuremath{\neq} entities under a heavy tail: a budget chosen by prior drift
preferentially catches high-frequency servers, so the event share
(\textasciitilde47\% at top-5\%) overstates how many changed servers are
actually re-checked. But entity coverage itself has a denominator choice
that must be stated. A history ranking can only rank servers that
existed at ranking time, so \emph{rankable-population coverage} - the
share of the \textbf{servers present at ranking time} that change and
are caught - is the measure of the policy\textquotesingle s performance
at its job. The whole-population view is lower, because roughly half of
all distinct servers that change in a test window are servers that
\emph{arrived after} the ranking cut and are outside any history
ranking\textquotesingle s reach by construction. We give both.

{\def\LTcaptype{none} 
\begin{longtable}[]{@{}lllllll@{}}
\toprule\noalign{}
Budget & 44d rankable cov. & 44d all-changer cov. & 44d event share &
60d rankable & 60d all-changer & 60d event \\
\midrule\noalign{}
\endhead
\bottomrule\noalign{}
\endlastfoot
top 5\% & \textbf{26.0\%} (5.2x) & 12.8\% & 47.5\% & \textbf{29.4\%}
(5.9x) & 16.7\% & 47.1\% \\
top 10\% & 39.1\% & 19.3\% & 60.5\% & 42.9\% & 24.4\% & 59.3\% \\
top 20\% & 53.6\% & 26.4\% & 71.2\% & 60.5\% & 34.4\% & 73.4\% \\
top 30\% & 61.3\% & 30.2\% & 74.6\% & 64.4\% & 36.6\% & 75.4\% \\
\end{longtable}
}

Read honestly: over the population it can rank, a top-5\% budget by
prior drift catches \textasciitilde26--29\% of the servers that will
change across the two cut-dates - a 5--6x lift over a same-size random
sample. Over \emph{all} changers it catches only
\textasciitilde13--17\%, because \textasciitilde43--51\% of test-window
changers are new arrivals it cannot rank, and §4.2 shows new servers are
exactly where drift concentrates. That blind spot is not a flaw in the
measurement but a structural limit of the policy, and it is a second
reason (alongside §5.3) that a history ranking is a partial control: it
can only cover servers it has already seen. The event share is not a
coverage number and we do not use it as one.

\subsubsection{5.3 The description surface resists
targeting}\label{53-the-description-surface-resists-targeting}

Repeating the identical protocol on description-only changes, the
top-5\% budget catches only \textbf{19.3\% / 21.5\%} of the
previously-seen servers whose description will change (rankable
coverage, \textasciitilde4x lift; \textasciitilde9--12\% of all
description changers), against 26--29\% for descriptor drift (5--6x);
the event-share figures are 25.3\% / 27.5\% versus 47\%. The gap is real
but smaller than the event share alone implied - the point is not that
description drift is unpredictable, but that a drift-history ranking,
tuned to the predictable version churn, buys measurably less on the
surface a description-level screen must recheck. Version churn is
habitual and publisher-driven; description rewrites are sparser, though
not markedly less predictable per server.

\textbf{Why the ranking caps, and why the top-5\% budget is the right
place to read it.} Measuring the conditional directly: prior description
change lifts next-period probability \textbf{4.8x} (13.8\% vs 2.8\%,
$n=3{,}492$), against 7.1x for descriptors. So the description ranking is
informative. What differs is how much of the population it can rank at
all. In the 44-day training window only \textbf{613 of 12,193 servers
(5.0\%)} carry any prior description change, against 2,000 (16.4\%) for
descriptors. A top-5\% budget is 609 slots and consumes essentially the
entire signal pool; taking the whole pool covers only \textbf{19.5\%} of
the rankable changers. Every slot beyond 5\% is therefore filled by
tie-break, not by signal - 49.7\% of a 10\% budget, 74.9\% of 20\%,
83.2\% of 30\%. This is what produces the otherwise anomalous
\emph{increasing} marginal returns of the description curve (+0.71,
+1.01, +1.78 coverage points per budget point across 5$\to$10$\to$20$\to$30):
the apparent improvement is not the ranking working better; past
exhaustion the slots carry no drift signal at all and are ordered by
publisher prefix, so what the curve measures there is that ordering. The shape beyond the 5\%
budget is therefore sort-order dependent, and visibly so: re-running the
deposited code on the released panel, whose tokenization permutes the
tie-break order, returns +0.86, +0.84, +0.69 instead. That the series
changes shape while the top-5\% figures move by at most 0.5 pp is the
clearest demonstration of the section's own claim - past the signal pool,
the curve measures tie-break order, not drift-ranking skill - and it is why we
read the result at the 5\% budget and nowhere else. The descriptor curve,
whose signal pool is not exhausted until a 16\% budget, shows the normal
decreasing shape (+2.63, +1.45, +0.77). The 60-day window replicates
both (description pool 829 / 5.9\%, ceiling 23.4\%). We therefore quote
the top-5\% figure not as a convenient budget but as the ranking's
ceiling: it is the point at which the method has spent everything it
knows. Ranking is deterministic - descending training-window change
count, ties broken by ascending server key - the tie-break therefore orders by publisher
prefix, and because change concentrates by publisher (§4.1) it is not
arbitrary with respect to future change - which is why we read coverage only
at the top-5\% budget, where the slots are signal-filled, and not at the
budgets past exhaustion. Running the deposited code on the released
(pseudonymized) panel, whose tokenization re-keys the tie-break, shows the
difference directly.\footnote{Top-5\% description coverage moves 19.3\%
\ensuremath{\rightarrow} 19.5\% between the named and released panels; top-30\%
moves 50.8\% \ensuremath{\rightarrow} 39.2\%. The deposit's
\texttt{DATASET\_CARD.md} tabulates the per-slot deltas.}

We consider this tension - not the headline
concentration - the paper\textquotesingle s most useful result
\textbf{for operators of description-level registry screens} (not for
tool-poisoning defense generally; §7.1), and it has a direct policy
implication: \textbf{do not rank by descriptor churn to catch
description change.} The control that fits the data is content-binding -
revalidate a verdict the moment its bound description hash moves -
backed by a periodic full-catalog sweep for the long tail and the new
arrivals a ranking cannot reach. Both controls address one failure mode - the judged text
changing after a verdict is minted. Whether the input was already stale when
the verdict was minted is a separate problem, which §6 illustrates but this
panel cannot size. Sizing the sweep concretely:
content-binding re-screens the \textasciitilde39 descriptions that change
per day (§3.3), while a sweep at cadence C days bounds worst-case
description staleness to C days at roughly N/C re-screens per day - for
the \textasciitilde19k-server corpus, \textasciitilde2.7k/day for a
weekly sweep or \textasciitilde630/day for a monthly one - a fixed,
plannable budget that a drift-history ranking\textquotesingle s
variable, blind-to-new-arrivals coverage does not give. (This is a
policy for \emph{description} revalidation only; the live-behavior
surface - the tool list and input schemas an attacker manipulates - is
not observed here, §7.1, and none of this bounds tool-poisoning
shelf-life.)

\subsubsection{5.4 A null result}\label{54-a-null-result}

Servers that received a public screening flag showed post-flag content
churn of 0.93x baseline (2,504 flagged server-days of exposure against
230,610) - no detectable change in
description-text churn after flagging. (This bounds text response only;
§7.1 applies.) Exposure
is small and flags may simply not be publisher-visible; we report it to
prevent its silent omission. Like the §6 counts, these derive from the
unreleased verdict store and are not recomputable from the deposited
panel.

\subsection{6. Consequences for a Deployed
Scanner}\label{6-consequences-for-a-deployed-scanner}

The scanner under study judges each server\textquotesingle s description
once at ingest and binds every verdict to \texttt{content\_hash}, the
SHA-256 of the exact text judged; until this work nothing re-compared
that binding to the registry. These counts come from the production
verdict store at the 2026-07-28 discovery snapshot; that store is not
released (§7), so these and the §5.4 flag-exposure counts are the
numbers a reader cannot recompute from the deposited panel. We report this as a deployment
observation, not a contribution.

\subsubsection{6.1 How stale were live
verdicts?}\label{61-how-stale-were-live-verdicts}

Joining all 18,543 live verdicts to the registry (99.6\% resolve):
\textbf{747 (4.03\% of all live verdicts) were bound to a description no longer published}
(an initial 781 included 34 whitespace-trap false positives, §2.3; the
panel corroborates 747 of 747 datable cases). Only 13 were also past
their clock-based expiry, so content-binding and wall-clock expiry are
nearly disjoint signals. A minority of the 747 carried a failing
dimension; we report that in aggregate, name no server, and publish no
per-record re-screen tally - a description hash cannot distinguish
remediation from rebranding from evasion, so the only defensible claim
is categorical: a point-in-time store can serve a failing verdict bound
to text no longer published, and here it did.

\subsubsection{6.2 Two producers of
staleness}\label{62-two-producers-of-staleness}

Dating the 747 against the panel splits them by producer: \textbf{379
(51\%) were born stale} - \texttt{evaluated\_at} postdates the last day
the judged text was published, the signature of the instrument reading a
lagged input - and \textbf{368 drifted after judgment}, judged against
then-current text a publisher later changed (379 + 368 = 747).
Post-remediation, drift runs at \textasciitilde26/day, against 34.6/day for
the panel\textquotesingle s independent description-change events over the
same period (§2.4) - the same order, with the deployment observing
somewhat fewer; a consistency check, not a precise agreement. Instrument lag - verdicts \emph{wrong at birth} rather than
aging - is invisible to any model that assumes the scanner sees the
present, is the larger share here, and is absent from the MCP threat
models we survey (§8).

\subsubsection{6.3 Mitigations}\label{63-mitigations}

The findings forced three production changes (live 2026-07-29). One
instantiates the §5.3 policy - read-time content-binding that renders a
mismatched verdict STALE without ever softening a failing verdict. The
other two address the separate failure mode §5.3 does not cover, staleness
at judgment time: a fail-closed input-freshness gate that refuses to judge a snapshot older
than 24 hours, so an input outage ages the store honestly instead of
minting born-stale verdicts; and a re-screen lane on the existing
catalog-sweep timer - the periodic sweep §5.3 sizes. That sweep also
backfilled \texttt{server\_id}, a subject-binding guard present in the
writer but unpopulated because no lane had ever re-run over judged
records - a reminder that a point-in-time pipeline can leave a schema
guard latent in the data until something forces a pass over the corpus.

\subsection{7. Limitations}\label{7-limitations}

\textbf{§7.1 Registry metadata, not probed behavior.} Descriptor drift
is not tool-behavior drift; a server can change its real tool list or
input schema - the surface an attacker ultimately manipulates - without
touching its registry entry, and our data never observes that surface.
This is the largest external-validity gap and the reason every defender
claim in this paper is scoped to \emph{description revalidation};
closing it (conformance probing of live tool lists) is ongoing work.

\textbf{§7.2 Self-reported tool counts} are regex-parsed prose from 861
of \textasciitilde19k servers.

\textbf{§7.3 Irregular cadence.} Observations land only when the
snapshot changed; intervals range from \textasciitilde2.5 hours to over
a day, and per-day normalization inherits that irregularity.

\textbf{§7.4 Single registry.} The largest prior audit spans six
registries {[}1{]}; our panel covers the official one. Generalization is
unproven.

\textbf{§7.5 Censoring.} "75.2\% never change" is right-censored at 88.6
days; the 89-day survival figure rests on nine cohorts drawn from the
small early corpus.

\textbf{§7.6 The scanner is our own.} §6 gains access no external
auditor would have, at the cost of studying one deployment; the
two-producer decomposition should be tested against other scanners, and
its counts are not independently replicable.

\subsection{8. Related Work}\label{8-related-work}

{[}1{]} audits 67,057 servers across six registries (collected late
June--early July 2025) and is, to our knowledge, the largest MCP
security measurement; its limitations discuss tool-extraction and
static-analysis precision but not temporal validity, though the authors
observed change directly ("After repeating the data collection\ldots{}
we identified only one expired GitHub token") without pursuing what
repetition implies. {[}2{]} studies 1,899 servers (March 2025) and
states the snapshot limitation explicitly - "Our study is limited to a
snapshot of MCP servers as of March 2025" - without quantifying it; this
paper is, in effect, the measurement that caveat calls for. {[}3{]}
analyzes 1,360 servers / 12,230 tools with no stated collection date and
no temporal discussion. {[}4{]} threat-models seven MCP clients with
tool poisoning as the central vector; its threats-to-validity section
covers scoring subjectivity and client generalization, and its
recommendations table calls for "periodic re-scanning" to catch tools
that "become malicious over time" - but it fixes no interval and
measures no rate. We are careful about the bridge here:
{[}4{]}\textquotesingle s target is the \emph{live tool}, which our data
does not observe (§7.1), so we do not supply its re-scan interval. What
we supply is a decay/revalidation rate for the \emph{registry
description text} a description-level screen reads;
{[}4{]}\textquotesingle s live-tool re-scan cadence remains unmeasured,
by us and by the prior work. Across all four, none reports a
revalidation interval or a decay rate for what it studies.

Longitudinal measurement of software-supply-chain ecosystems is well
established outside MCP: for the npm package graph and its security
threats {[}5{]}, for the corpus of open-source supply-chain attacks
{[}6{]}, and for the web-PKI certificate ecosystem {[}7{]}. Our
contribution is bringing that lens to MCP: the first longitudinal panel
of an MCP registry, and the finding that drift-history re-auditing
under-covers the description-revalidation surface (§5.3), with the
content-binding-plus-sweep policy that follows. The §6 instrument-lag
observation is a deployment note, not a contribution claim - it comes
from a single unreleased store and is not independently replicable -
though it does illustrate a staleness component invisible without paired
scanner-side data.

\subsection{Data and Code
Availability}\label{data-and-code-availability}

The pseudonymized MCP Registry Drift Panel v1 and the full analysis code
are released under CC-BY-4.0. Every panel-derived number in this paper
regenerates from the deposit, with four stated exceptions. First, the publisher-level concentration in
§4.1, which needs the plaintext publisher prefix that pseudonymization
destroys (it is marked \texttt{needs\_named\_panel} in
\texttt{figures.json}). Second, the §3.2 version-recency estimator, which
reads \texttt{publishedAt} from the registry snapshot rather than the panel.
Third, the §5.4 and §6 counts, which come from the unreleased verdict store.
Fourth, budget slots filled by tie-break rather than signal: the ranking
sorts by (-change count, server key), and pseudonymization permutes the key
space, so targeting figures beyond the exhausted signal pool select
different servers in the released panel than in the named one - at most
0.5 pp at the top-5\% budget where every headline claim lives, up to ~12 pp
at a 30\% budget, with the per-slot deltas tabulated in the deposit's
\texttt{DATASET\_CARD.md}. Otherwise:
\texttt{paper\_figures.py} produces
\texttt{figures.json}, \texttt{paper\_figures\_addenda.py} produces
\texttt{figures\_addenda.json} (§4.2, §4.4, §5.1, §5.3), and
\texttt{make\_paper\_figure.py} renders Figure 1. The first two run against
the deposited panel with no arguments and no dependencies beyond the
Python standard library; the figure renderer additionally requires
\texttt{matplotlib}. The panel is deposited at Zenodo (concept DOI
10.5281/zenodo.21709945, which resolves to the current version) and
pinned to the official MCP registry snapshot captured at commit
\texttt{e1d19d3} (data cutoff 2026-07-28; 120 observations); the exact
bytes of the released panel are fixed by \texttt{SHA256SUMS} in the deposit
and by the version DOI 10.5281/zenodo.21798111
(\texttt{figures.json:\_pin.panel\_sha256} is the digest of the internal
named panel the figures were computed from, not of the released file). Two earlier datasets from this project are also cited:
MCP Drift v1 (DOI 10.5281/zenodo.21449150) and Source Liveness v1 (DOI
10.5281/zenodo.21501868). The production verdict store underlying §5.4 and §6 is
a single unreleased deployment (§7).

\subsection*{Competing interests}

The scanner described in §6 is mcpindex.ai, the
author\textquotesingle s own production system and a commercial service. The §5.3 recommendation is derived from
§§3--5 on the deposited panel and does not depend on §6; §6 is reported as
a deployment observation and is excluded from the paper\textquotesingle s
contribution claims (§1, §5.4, §7.6).

\subsection{References}\label{references}

{[}1{]} Xiaofan Li and Xing Gao. "A First Look at the Security Issues in
the Model Context Protocol Ecosystem." arXiv:2510.16558, 2025.

{[}2{]} Mohammed Mehedi Hasan, Hao Li, Emad Fallahzadeh, Gopi Krishnan
Rajbahadur, Bram Adams, and Ahmed E. Hassan. "Model Context Protocol
(MCP) at First Glance: Studying the Security and Maintainability of MCP
Servers." arXiv:2506.13538, 2025.

{[}3{]} Shuli Zhao, Qinsheng Hou, Zihan Zhan, Yanhao Wang, Yuchong Xie,
Yu Guo, Libo Chen, Shenghong Li, and Zhi Xue. "Parasites in the
Toolchain: A Large-Scale Analysis of Attacks on the MCP Ecosystem."
arXiv:2509.06572, 2025.

{[}4{]} Charoes Huang, Xin Huang, Ngoc Phu Tran, and Amin Milani Fard.
"Model Context Protocol Threat Modeling and Analysis of Vulnerabilities
to Prompt Injection with Tool Poisoning." Journal of Cybersecurity and
Privacy 6(3):84, 2026. DOI: 10.3390/jcp6030084. (Preprint:
arXiv:2603.22489.)

{[}5{]} Markus Zimmermann, Cristian-Alexandru Staicu, Cam Tenny, and
Michael Pradel. "Small World with High Risks: A Study of Security
Threats in the npm Ecosystem." USENIX Security Symposium, 2019.

{[}6{]} Marc Ohm, Henrik Plate, Arnold Sykosch, and Michael Meier.
"Backstabber\textquotesingle s Knife Collection: A Review of Open Source
Software Supply Chain Attacks." DIMVA, 2020.

{[}7{]} Zakir Durumeric, James Kasten, Michael Bailey, and J. Alex
Halderman. "Analysis of the HTTPS Certificate Ecosystem." Internet
Measurement Conference (IMC), 2013.

\end{document}